\documentclass[a4paper,10pt,twoside]{cpc-hepnp}

\usepackage{multicol}
\usepackage{graphicx}
\usepackage{booktabs}
\usepackage{amssymb,bm,mathrsfs,bbm,amscd}
\usepackage[tbtags]{amsmath}
\usepackage{lastpage}

\begin{document}

\fancyhead[co]{\footnotesize SUN Le-Xue~ et al}


\title{Hadronic Rapidity Spectra in Heavy Ion Collisions at SPS and AGS energies in a Quark Combination Model\thanks{Supported by National Natural Science
Foundation of China (10775089,10947007,10975092) }}

\author{%
      SUN Le-Xue$^{1}$
\quad WANG Rui-Qin$^{2}$
\quad SONG Jun$^{2}$
\quad SHAO Feng-Lan$^{1;1)}$\email{shaofl@mail.sdu.edu.cn}%
}
\maketitle

\address{%
$^1$ Department of Physics, Qufu Normal University, Shandong 273165, China\\
$^2$ Department of Physics, Shandong University, Shandong 250100, China\\
}

\begin{abstract}
The quark combination mechanism of hadron production is applied to nucleus-nucleus collisions at the CERN Super Proton Synchrotron (SPS) and BNL Alternating Gradient Synchrotron (AGS).
The rapidity spectra of identified hadrons and their spectrum widths are studied.
The data of $\pi^{-}$, $K^{\pm}$, $\phi$, $\Lambda$, $\overline{\Lambda}$, $\Xi^{-}$, and $\overline{\Xi}^{+}$ at 80 and 40 AGeV, in particular at 30 and 20 AGeV where the onset of deconfinement is suggested to happen, are consistently described by the quark combination model. However at AGS 11.6 AGeV below the onset the spectra of $\pi^{\pm}$, $K^{\pm}$ and $\Lambda$ can not be simultaneously explained, indicating the disappearance of intrinsic correlation of their production in the constituent quark level.
The collision-energy dependence of the rapidity spectrum widths of constituent quarks and the strangeness of the hot and dense quark matter produced in heavy ion collisions are obtained and discussed.

\end{abstract}



\begin{multicols}{2}

\section{Introduction}

The production of quark gluon plasma (QGP) and its properties are hot topics in relativistic heavy ion collisions.
A huge number of possible QGP signals were proposed and measured, and many unexpected novel phenomena were observed
at RHIC and SPS \cite{star_review,phenix_review,brahms_review,phobos_review,NA49_onset}. These experimental data
greatly contribute to the identification of QGP and the understanding of its properties and hadronization
from different aspects. Especially, there are a class of phenomena that are of particular interest,
i.e. the abnormally high
ratio of baryons to mesons and the quark number scaling of hadron elliptic flows in the intermediate $p_{T}$ range, etc\cite{phenixv2,phenixbmratio}. They reveal the novel features of hadron production
in relativistic heavy ion collisions.

In quark combination/coalescence scenario, hadrons are combined from quarks and antiquarks, i.e., a quark-antiquark pair merges into a meson and three quarks into a baryon. The production difference between baryon and meson mainly results from their different constituent quark numbers. It is shown that such a simple quark number counting can naturally explain those
striking features of hadron production observed at RHIC\cite{Greco2003PRL,Fries2003PRL,v2highpt},
while the fragmentation mechanism can not.

The highlights at RHIC are mainly of hadron production in transverse direction where the quark combination scenario mostly flashes. In fact,
the longitudinal rapidity distribution of hadrons is also a
good tool for testing the hadronization mechanism. In previous work\cite{JSong2009MPA,CEShao2009PRC}, we have used the quark combination
model to successfully describe the rapidity spectra of various hadrons at RHIC $\sqrt{s_{NN}}=200$ GeV and top SPS $E_{beam}=158$ AGeV. At other collision energies where the QGP may be produced, e.g., at lower SPS and higher LHC
energies, does the quark combination mechanism still work well?
The Beam Energy Scan at RHIC and the NA49 Collaboration have provided
abundant data on hadron production in the energy region from 20 GeV to 6 GeV.
In this paper, we extend the quark combination model to systematically study the rapidity distributions of various identified hadrons in heavy ion collisions at SPS $E_{beam}=80,~40,~30,~20$ AGeV and AGS $E_{beam}=11.6$ AGeV and test the applicability of the quark combination mechanism.

\section{A brief introduction to the quark combination model}

The quark combination model deals with how quarks and antiquarks covert to color-siglet hadrons
as the partonic matter evolves to the interface of hadronization. The basic idea
is to put all the quarks and antiquarks line up in a one-dimensional
order in phase space, e.g., in rapidity, and let them combine
into primary hadrons one by one following a combination rule based on the
QCD and near-correlation in phase space requirements.
See Sec. II of Ref.\cite{CEShao2009PRC} for detailed description of such a rule. Here, we consider only
the production of SU(3) ground states, i.e. 36-plets of mesons and 56-plets of baryons. The flavor
SU(3) symmetry with strangeness suppression in the yields
of initially produced hadrons is fulfilled in the model.
The decay of short-life hadrons is systematically taken into account
to get the spectra comparable to the data.
The model has reproduced the experimental
data for hadron multiplicity ratios, momentum distributions and the elliptic
flows of identified hadrons, etc., in
heavy ion collisions at
RHIC and top SPS energies\cite{FLShao2005PRC,FLShao2007PRC,YFWang2008CPC,Yao08prc,CEShao2009PRC}, and
addressed the entropy issue\cite{sjEntropy} and the exotic state production\cite{WHan2009PRC}.

\section{Rapidity spectra of constituent quarks }

The rapidity spectra of constituent quarks just before hadronization are needed as the input of the quark combination model.
Considering that the collision energies studied here are much lower than RHIC energies, and in particular 30-20 AGeV is the possible region for the onset of deconfinement, it is not sure whether the hot and dense quark matter is exactly produced
at these energies, so applying a model or theory for the evolution of the hot and dense quark matter, e.g., relativistic
hydrodynamics, to get the quark spectra before hadronization may be uneconomic or infeasible.
In this paper, assuming the hot quark matter has been created, we parameterize the rapidity distribution of constituent quarks and extract the parameter values from the experimental data of final state hadrons.
The rapidity distribution of newborn quarks is taken to be a Gaussian-like form, i.e.,
\begin{equation}
\frac{dN_q}{dy}=N_q f_{q}(y)= \frac{N_{q} }{A} \Big( e^{-|y|^{a}/2\sigma^{2}} -C \Big).
\end{equation}
Here $C=\exp[{-|y_{beam}|^{a}/2\sigma^{2}}]$ which means the constituent quarks that form hadrons via combination are within $[-y_{beam}, y_{beam}]$ in the center-of-mass frame. $A$ is the normalization factor, satisfying $A=\int^{y_{beam}}_{-y_{beam}} \big (e^{-|y|^{a}/2\sigma^{2}} -C \big) dy$. $a$ and $\sigma$ are the parameters depicting the shape of the spectrum. $N_q$ denotes the total number of newborn quarks with type $q$ in the full rapidity region.
For the net quarks coming from the colliding nuclei, their evolution is generally different from the newborn quarks due to complicated collision transparency \cite{bearden04stop}. We fix the rapidity spectrum of net-quarks before hadronization by the data of the rapidity distribution of (net-)protons\cite{2080proton}, and the results are shown in Fig. \ref{netq-y}.

\begin{center}
  \includegraphics[width=0.6\linewidth]{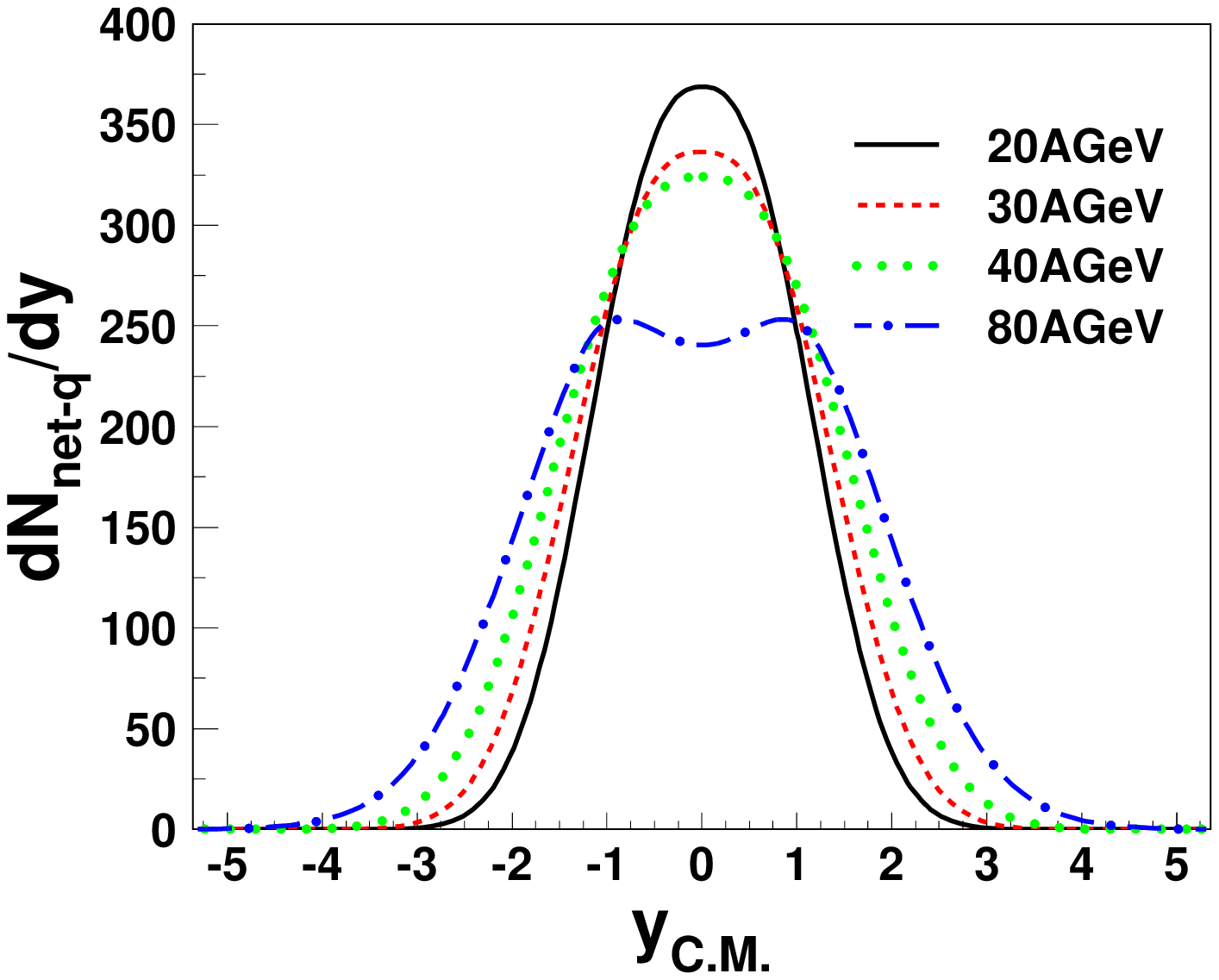}\\
  \figcaption{The rapidity distributions of net-quarks.}\label{netq-y}
\end{center}

\section{Hadronic rapidity spectra at SPS and AGS}

In this section, we first use the quark combination model to systematically study hadron rapidity spectra and their widths at SPS energies. Then we present the energy dependence of the strangeness and spectrum width of constituent quarks. Finally, we show the results at AGS 11.6AGeV.

\subsection{Rapidity spectra of hadrons at SPS energies}
\end{multicols}
\ruleup

\begin{center}
  \includegraphics[width=0.8\linewidth]{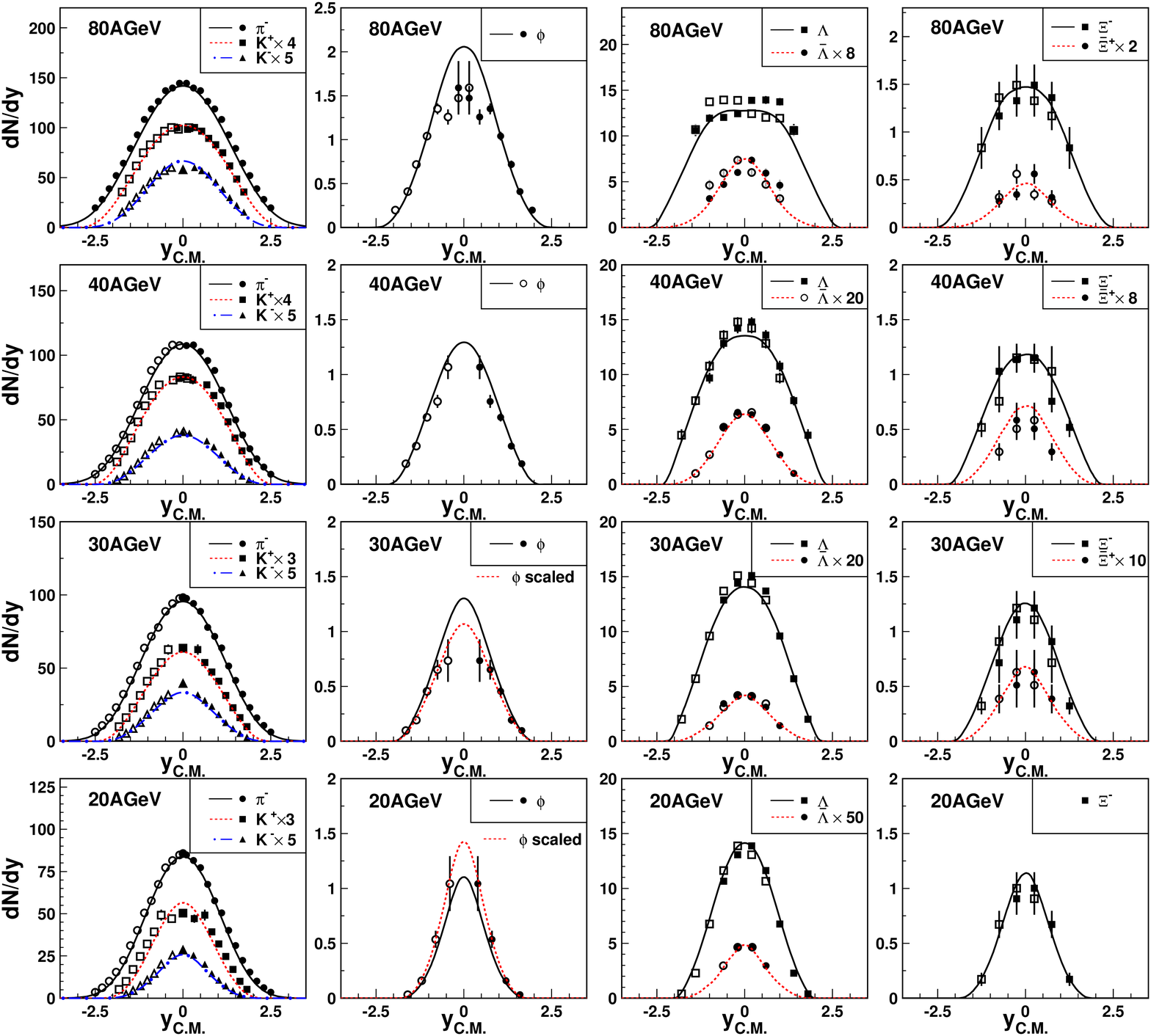}\\
  \figcaption{Rapidity distributions of identified hadrons in central Pb+Pb collisions at $E_{beam} =~80,~40,~30,~20$ AGeV. The symbols are the experimental data\cite{158A2002PRC,Afanasiev2000Phi,Alt08Xi,Mischke03Lam,phiEDNA49} and lines are the calculated results. The open symbols are reflection of measured data (solid symbols) against midrapidity.
  The results of $\phi$ at 30, 20 AGeV are scaled by proper constant factors for comparing their shapes with the data.}
  \label{sps}
\end{center}
\ruledown
\begin{multicols}{2}

The calculated rapidity distributions of identified hadrons in central Pb+Pb collisions at $E_{beam} =$ 80, 40, 30, 20 AGeV are shown in Fig. \ref{sps}.  The values of parameters for newborn quarks are listed in Table \ref{qpars}, and $\chi^{2}/ndf$ is presented as the quality of global fitting. Note that the experimental data of pions beyond $y_{beam}$ are not included
in $\chi^{2}/ndf$ calculation, because the limiting fragmentation behavior around $y_{beam}$ becomes prominent which is beyond the study of this paper by quark combination.
The results show the quark combination model can reproduce the experimental data of
various identified hadrons except $\phi$ at 30, 20 AGeV.
The yields of $\phi$ at 30, 20 AGeV deviate from the data, but their sharps are still in agreement with the data after
scaling by proper constant factors.
\end{multicols}
\ruleup
\begin{center}
\tabcaption{Parameters of newborn light and strange quarks and $\chi^{2}/ndf$.}
{\begin{tabular*}{0.7\linewidth}{@{\extracolsep\fill}cccccccccc}
\toprule
energy & $a_{u/d}$ & $a_{s}$ & $\sigma_{u/d}$ & $\sigma_{s}$ & $N_{u/d}$ & $N_{s}$ & $\chi^{2}/ndf$ \\
\hline
80AGeV    &1.90  &2.35   &1.19   &1.40   &275.2    &159.6   &0.98  \\
40AGeV    &1.90  &2.25   &1.12   &1.40   &171.3    &116.5   &1.01  \\
30AGeV    &1.80  &2.05   &1.20   &1.16   &139.0    &101.5   &0.96  \\
20AGeV    &1.80  &1.95   &1.10   &0.85   &105.5    &77.0    &1.90  \\
\bottomrule
\end{tabular*} \label{qpars}}
\end{center}
\vspace{6mm}
\ruledown
\begin{multicols}{2}

The deviation of $\phi$ yields at 30 and 20 AGeV is related to the pronounced rescattering effect. It is shown that at higher SPS and RHIC energies the production of $\phi$ mainly comes from the contribution of partonic phase, i.e., the directly produced $\phi$ by hadronization while at lower SPS energies kaon coalescence may be the dominated production mechanism\cite{phiEDNA49}. In addition, the directly produced $\phi$ after hadronization will possibly suffer the destroy by the scattering of the daughter kaons with other produced hadrons. The absence of these two effects at hadronic stage in our calculations is the main reason for the deviation in $\phi$ yields.

\subsection{Widths of rapidity spectra for hadrons}

From the above experimental data one can see the widths of rapidity spectra for different hadron species are generally different. This difference, in other word the correlation of longitudinal hadron production, can be used to quantitatively test various hadronization models. In quark combination mechanism, the rapidity distribution of a specific hadron is the convolution of the rapidity spectra of its constituent quarks and combination probability function (denoted by the combination rule in our model). Since the rapidity distributions of different-flavor quarks are different (see Table \ref{qpars} and Fig. \ref{netq-y}), in particular the difference between newborn quarks and net-quarks, the shapes of rapidity spectra of hadrons with different quark flavor components are generally different and correlated with each other by constituent quarks. Here, we calculate the spectrum widths of various hadrons to clarify this feature.

Considering that the rapidity region covered by the data for different hadron species or at different collision energies are not the same, we define the variance of rapidity distribution for a specified hadron in a finite rapidity region limited by the discrete experimental data,
\begin{equation}
  \langle y^{2} \rangle ^{(L)}=\frac{\sum\limits_{i=1}^{n} y_{i}^{2}\frac{dN_{i}}{dy}}{\sum\limits_{i=1}^{n} \frac{dN_{i}}{dy}}.
\end{equation}
Here $n$ is the number of experimental data; $y_{i}$ and $dN_{i}/dy$ are the rapidity position and the corresponding yield density measured experimentally, respectively. Replacing $dN_{i}/dy$ with the model results, we can give the $\langle y^{2} \rangle ^{(L)}|_{model}$ and compare it with the experimental value $\langle y^{2} \rangle ^{(L)}|_{data}$ to test the applicability of the model without any arbitrariness caused by the rapidity region where the experimental data have not covered yet. We further extrapolate $\langle y^{2} \rangle ^{(L)}|_{data}$ to the full rapidity region $[-y_{beam},y_{beam}]$ via the relation
\begin{equation}
  \langle y^{2} \rangle ^{(F)}|_{data}=\frac{\langle y^{2} \rangle ^{(L)}|_{data}}{\langle y^{2} \rangle ^{(L)}|_{model}} \langle y^{2} \rangle ^{(F)}|_{model},
\end{equation}
where $\langle y^{2} \rangle ^{(F)}|_{model}$ is the variance calculated by the model in the full rapidity region. The degree of agreement between $\langle y^{2} \rangle ^{(F)}|_{model}$ and experimental data $\langle y^{2} \rangle ^{(F)}|_{data}$ still represents the original description ability of the model, and the $<y^2>$ of different hadron species can be directly compared and their energy dependence is recovered.
The spectrum widths of various hadrons, i.e. $D(y)^{(F)}\equiv\sqrt{\langle y^{2} \rangle ^{(F)}}$, are calculated and the results are shown in Fig \ref{avey2} (158AGeV is also included). One can see that the $D(y)$ of various hadrons given by the model are obviously distinguished.  The agreement between the data and the calculated results is the support of the theoretic(quark recombination mechanism) explanation for the widths of hadronic rapidity distributions.
\end{multicols}
\ruleup

\begin{center}
  \includegraphics[width=0.7\linewidth]{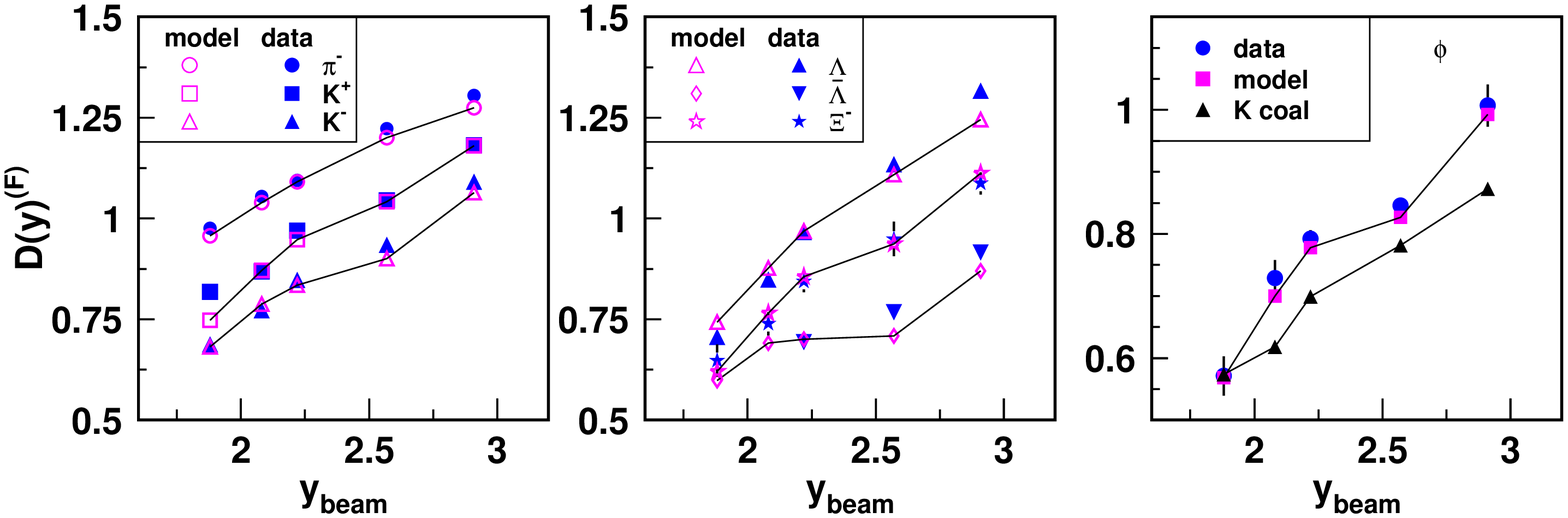}\\
  \figcaption{Widths of rapidity distributions $D(y)^{(F)}$ at $E_{beam}(y_{beam})=158,~80,~40,~30,~20$ AGeV (2.91, 2.57, 2.22, 2.08, 1.88). The right panel includes the results of $K^{+}K^{-}$ coalescence for $\phi$ production. Model results are connected by the solid lines to guide the eye.}\label{avey2}
\end{center}
\ruledown

\begin{multicols}{2}

As stated above, at lower SPS energies, kaon meson coalescence may be an important mechanism for $\phi$ production. The right panel of Fig. \ref{avey2} shows the $\phi$'s $D(y)^{(F)}|_{data}$ and $D(y)^{(F)}|_{model}$ as well as the results for $K^{+} K^{-}$ coalescence $D(y)^{(F)}|_{coal}$. Similar to Ref.\cite{phiEDNA49}, considering the ideal case of coalescence of two kaons with the same rapidity, we use the measured kaon rapidity distributions $f_{K^{\pm}}(y)$ to obtain the spectrum of $\phi$ by $f_{\phi}^{coal}(y) \propto f_{K^{+}}(y) f_{K^{-}}(y) $ and then give the $D(y)^{(F)}|_{coal}$. One can see that at collision energies above 20 AGeV, $D(y)^{(F)}|_{coal}$ is much smaller than the data, which is similar to the results in Ref.\cite{phiEDNA49}, while our results nearly agree with the data. This clearly shows the $\phi$ production at these energies is dominated by the hadronization. At 20 AGeV, $D(y)^{(F)}|_{coal}$ is nearly equal to the data and the model result is also in agreement with the data. It suggests $\phi$ production at lowest SPS energy can have different explanations, in other words, even though kaon coalescence is significant, the $\phi$ directly produced from hadronization may be unnegligible.

\subsection{The strangeness and spectrum width of constituent quarks}

Let us turn to the extracted rapidity distributions of newborn light and strange quarks. Their properties can be characterized by two quantities, i.e the ratio of strange quark number $N_s$ to light quark number $N_{u/d}$ called strangeness suppression factor $\lambda_{s}$ and the width of rapidity spectrum.

The left panel of fig. \ref{parton} shows the strangeness suppression factor $\lambda_{s}$ at different collision energies. Note that $\lambda_{s}$ defined here is in terms of quark numbers in the full phase space, so the values are sightly different from those in terms of midrapidity quark number densities in previous Ref. \cite{CEShao2009PRC}.  As the comparison with SPS, we also present $\lambda_{s}$ at RHIC $\sqrt{s_{NN}}=200,~62.4$ GeV calculated by the model and at AGS 11.6 AGeV calculated by counting the numbers of light and strange valance quarks hidden in the observed pions, kaons and $\Lambda$ which are mostly abundant hadron species carrying light and strange gradients of the system. One can see that the value of $\lambda_{s}$ exhibits a peak behavior at lower SPS energies. This behavior has been reported by NA49 Collaboration as the signal of onset of deconfinement.

\end{multicols}
\ruleup
\begin{center}
  \includegraphics[width=0.6\linewidth]{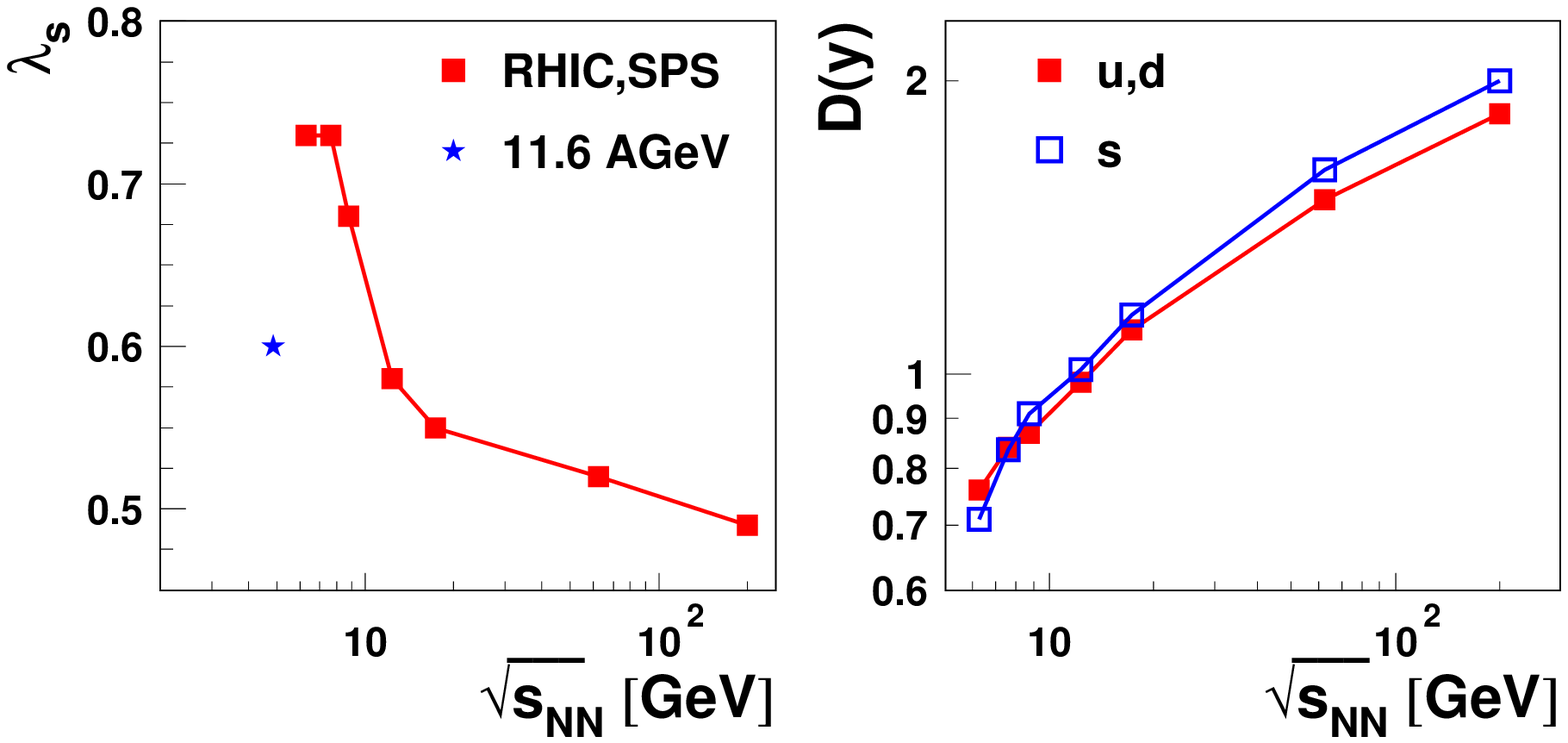}\\
  \figcaption{The strangeness suppression factor $\lambda_{s}$ and $D(y)$ of newborn light and strange quarks at different collision energies. The results are connected by the solid lines to guide the eye. }
  \label{parton}
\end{center}
\ruledown

\begin{multicols}{2}
The width of rapidity spectrum is characterized by $D(y)\equiv\sqrt{\langle y^2\rangle -\langle y\rangle^2}=\sqrt{\langle y^2\rangle}$. The right panel of Fig. \ref{parton} shows the $D(y)$ of newborn light and strange quarks at different collision energies. The results at RHIC $\sqrt{s_{NN}}=200,~62.4$ GeV and SPS $E_{beam}=158$ AGeV are included.
One can see that with the increasing of collision energy $D(y)$ of newborn light and strange quarks both increase regularly, indicating the stronger collective flow formed at higher energies.
As collision energy is equal to or greater than 40 AGeV, $D(y)$ of newborn light quarks are always smaller than that of strange quarks while at 30 and 20 AGeV the situation flips.
The wider spectrum of strange quarks, relative to light quarks, has been verified at RHIC both in the longitudinal and transverse directions\cite{CEShao2009PRC,YFWang2008CPC,ChenJH2008}, and the explanation is that strange quarks acquire stronger collective flow during evolution in partonic phase.
As the collisions energy reduces to 30 and 20 AGeV, the widths of rapidity distributions of light and strange quarks, see Table \ref{qpars} and Fig. \ref{parton}, are all quite narrow,  which means the collective flow formed in partonic phase is much smaller than those at higher SPS and RHIC energies.
Therefore,  the partonic bulk matter,  even if produced as the indication of our results via still active constituent quark degrees of freedom, should be in the vicinity of phase boundary, and the extracted momentum distributions of quarks keep the memory of their original excitation.
If thermal equilibrium is reached in heavy ion collisions, the quark occupation function in momentum space follows as $\exp[-E/T]=\exp[-m_{T} \cosh(y)/T]$ in the case of no collective flow.  Taking the hadronization temperature $T=165$ MeV and constituent mass $m_{q}=340$ MeV for light quarks and $m_{s}=500$ MeV for strange quarks, the quark rapidity spectrum is Gaussian form and width of light quarks is $\sigma_q=0.6$ and strange quarks $\sigma_s=0.52$ due to the heavier mass.  The tighter spread of strange quarks in rapidity space can be qualitatively understood in quark production with thermal-like excitation.

\subsection{Results at AGS 11.6 AGeV}

What happens at lower AGS energies? We further use the model to calculate the rapidity distributions of various hadrons at 11.6 AGeV. The results are shown in Fig. \ref{ags} and are compared with the experimental data. The values of parameters ($a, \sigma, N_q$) for quark spectra are taken to be (2.1, 0.88,71) for newborn light quarks and (2.0, 0.83, 42) for strange quarks, respectively. The rapidity distribution of net-quarks is extracted from the proton data of  E802 Collaboration \cite{E802piKP,E802PionPmid}, and the data of E877 Collaboration \cite{E877piPForward} at 10.8 AGeV are used as the guide of extrapolation of net-quark spectrum in the forward rapidity region. One can see that the results for pions and kaons are in agreement with the data but the result of $\Lambda$ can not reproduce the experimental data --- the width of spectrum given by the model is much wider than that of data.
This suggests that there is no intrinsic correlation at constituent quark level between production of kaons and $\Lambda$ at AGS 11.6 AGeV.
In addition, the rapidity distributions of $\phi$, $K_{s}^{0}$, $\overline{p}$, $\overline{\Lambda}$ and $\Xi^{-}(\overline{\Xi}^{+})$ are predicted
to be tested by future data.
\end{multicols}
\ruleup
\begin{center}
  \includegraphics[width=0.7\linewidth]{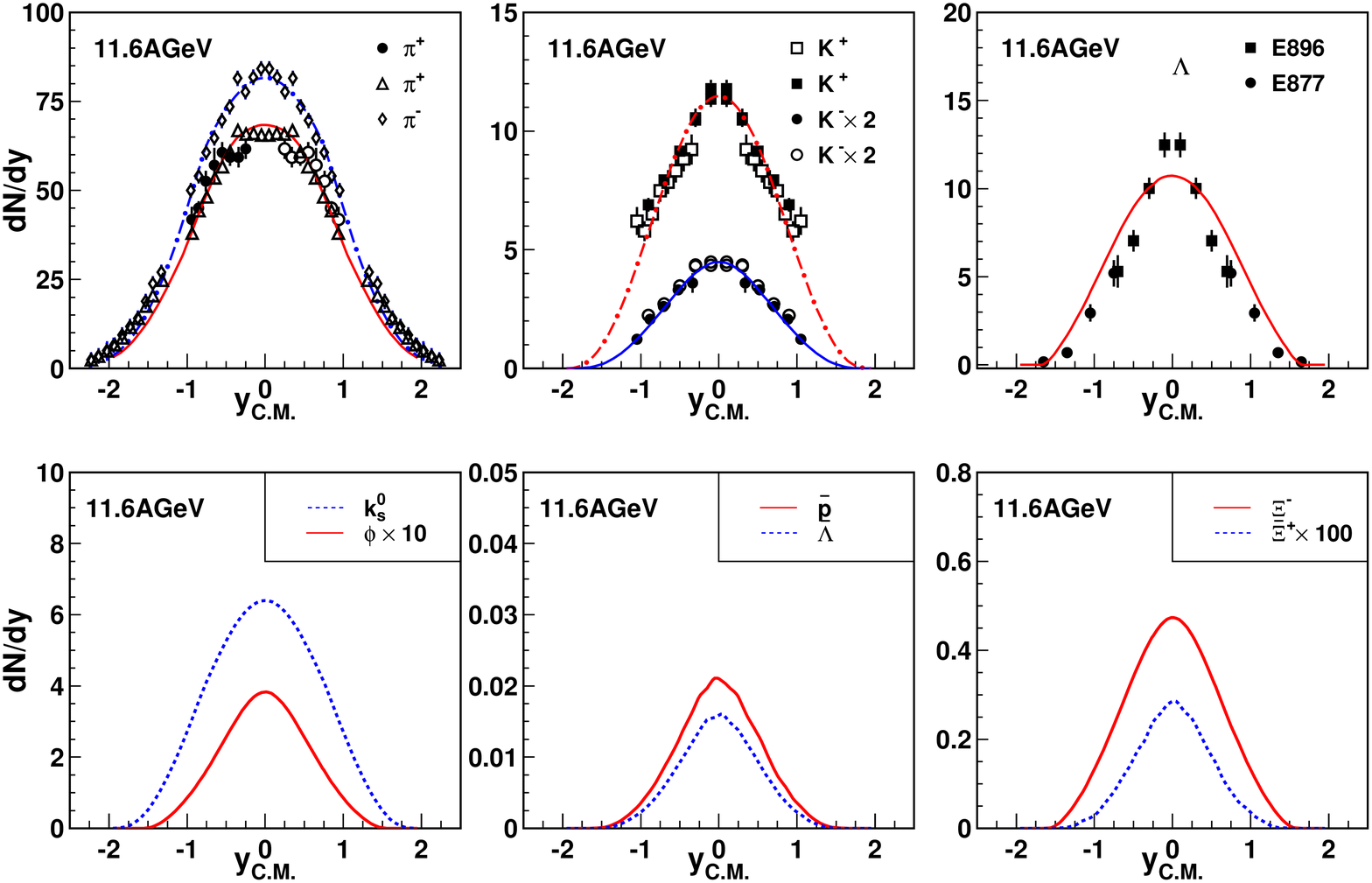}\\
  \figcaption{Rapidity distributions of identified hadrons in central Au+Au collisions at $E_{beam} =$ 11.6 AGeV. The symbols are experimental data from Refs \cite{E877piPForward,E802piKP,E802K98,E802PionPmid,E896Lam,E877Lam} and lines are the calculated results.}
  \label{ags}
\end{center}
\ruledown
\begin{multicols}{2}

\section{Summary}
In this paper, we have investigated, with the quark combination model, the rapidity distributions of identified hadrons and their widths in central A+A collisions at SPS and AGS energies.
Assuming in advance the existence of constituent quark degrees of freedom, we parameterize the rapidity spectra of quarks before hadronization, then test whether such a set of light and strange quark spectra can self-consistently explain the data of  $\pi^{-}$, $K^{\pm}$, $\phi$, $\Lambda(\overline{\Lambda})$, $\Xi^{-}(\overline{\Xi}^{+})$, etc., at these energies.
The results of hadronic rapidity spectra are in agreement with the data at 80 and 40 AGeV. At 30 and 20 AGeV where the onset of deconfinement is suggested to happen, the model can still basically describe the production of various hadrons.
The study of rapidity-spectrum widths for hadrons, particularly for phi via the comparison to the results from kaon coalescence in the stage of hadronic rescatterings, clearly show the hadron production at the collision energies above 20 AGeV is dominated by the hadronization.
The energy dependence of the rapidity-spectrum widths of constituent quarks and the strangeness of the hot and dense quark matter are obtained. It is shown that the strangeness peaks around 30 AGeV and below (above) the energy, the width of strange quarks becomes to be smaller (greater) than that of light quarks.
As the collision energy decreases to AGS 11.6 AGeV, it is found that the production of $\pi^{\pm}$, $K^{\pm}$ and $\Lambda$ can not be consistently explained by the model, which suggests there seems to be no intrinsic correlations for their production in the constituent quark level.
If the applicability of the quark combination mechanism can be regarded as a judgment of the QGP creation, our results implies the threshold for the onset of deconfinement is located in the energy region 11.6-20 AGeV, which is consistent with the report of NA49 Collaboration.

\acknowledgments{The authors would like to thank professor XIE Qu-Bing for fruitful discussions.}

\end{multicols}


\begin{multicols}{2}

\end{multicols}

\clearpage

\end{document}